\documentclass[twocolumn,nofootinbib,preprintnumbers,superscriptaddress,amsmath,amssymb,PRL]{revtex4}
\usepackage{graphicx}
\usepackage{dcolumn}
\usepackage{float}
\usepackage{mathptmx, courier, pifont}
\usepackage[scaled=0.92]{helvet}
\usepackage[T1]{fontenc}
\usepackage{textcomp}
\usepackage{color}
\usepackage[normalem]{ulem}

\usepackage[dvipsnames]{xcolor}
\usepackage[colorlinks=true,urlcolor=Blue,linkcolor=Blue]{hyperref}
\usepackage[all]{hypcap}
\usepackage{amsmath}

\begin{document}


\title{Masses of neutron-rich $^{\operatorname{52-54}}$Sc and $^{54,56}$Ti nuclides: The $N=32$ subshell closure in scandium}

\author{X.~Xu}
\affiliation{Key Laboratory of High Precision Nuclear Spectroscopy and Center for Nuclear Matter Science, Institute of Modern Physics, Chinese Academy of Sciences, Lanzhou 730000, People's Republic of China}
\affiliation{School of Science, Xi$^{\prime}$an Jiaotong University, Xi$^{\prime}$an, 710049, China}
\author{M.~Wang}\thanks{Corresponding author: wangm@impcas.ac.cn}
\affiliation{Key Laboratory of High Precision Nuclear Spectroscopy and Center for Nuclear Matter Science, Institute of Modern Physics, Chinese Academy of Sciences, Lanzhou 730000, People's Republic of China}
\author{K.~Blaum}
\affiliation{Max-Planck-Institut f\"{u}r Kernphysik, 69117 Heidelberg, Germany}
\author{J.~D.~Holt}
\affiliation{TRIUMF, 4004 Wesbrook Mall, Vancouver, British Columbia V6T 2A3, Canada}
\author{Yu.~A. Litvinov}\thanks{Corresponding author: y.litvinov@gsi.de}
\affiliation{Key Laboratory of High Precision Nuclear Spectroscopy and Center for Nuclear Matter Science, Institute of Modern Physics, Chinese Academy of Sciences, Lanzhou 730000, People's Republic of China}
\affiliation{GSI Helmholtzzentrum f\"{u}r Schwerionenforschung, 64291 Darmstadt, Germany}
\author{A.~Schwenk}
\affiliation{Institut f\"ur Kernphysik, Technische Universit\"at Darmstadt, 64289 Darmstadt, Germany}
\affiliation{ExtreMe Matter Institute EMMI, GSI Helmholtzzentrum f\"ur Schwerionenforschung GmbH, 64291 Darmstadt, Germany}
\affiliation{Max-Planck-Institut f\"{u}r Kernphysik, 69117 Heidelberg, Germany}
\author{J.~Simonis}
\affiliation{Institut f\"{u}r Kernphysik and PRISMA Cluster of Excellence, Johannes Gutenberg-Universit\"at, 55099 Mainz, Germany}
\affiliation{Institut f\"ur Kernphysik, Technische Universit\"at Darmstadt, 64289 Darmstadt, Germany} 
\affiliation{ExtreMe Matter Institute EMMI, GSI Helmholtzzentrum f\"ur Schwerionenforschung GmbH, 64291 Darmstadt, Germany} 
\author{S.~R.~Stroberg}
\affiliation{TRIUMF, 4004 Wesbrook Mall, Vancouver, British Columbia V6T 2A3, Canada}
\affiliation{Reed College, Portland, 97202 Oregon, USA}
\author{Y.~H.~Zhang}
\affiliation{Key Laboratory of High Precision Nuclear Spectroscopy and Center for Nuclear Matter Science, Institute of Modern Physics, Chinese Academy of Sciences, Lanzhou 730000, People's Republic of China}
\affiliation{ExtreMe Matter Institute EMMI, GSI Helmholtzzentrum f\"ur Schwerionenforschung GmbH, 64291 Darmstadt, Germany} 
\author{H.~S.~Xu}
\affiliation{Key Laboratory of High Precision Nuclear Spectroscopy and Center for Nuclear Matter Science, Institute of Modern Physics, Chinese Academy of Sciences, Lanzhou 730000, People's Republic of China}
\author{P.~Shuai}
\affiliation{Key Laboratory of High Precision Nuclear Spectroscopy and Center for Nuclear Matter Science, Institute of Modern Physics, Chinese Academy of Sciences, Lanzhou 730000, People's Republic of China}
\author{X.~L.~Tu}
\affiliation{Key Laboratory of High Precision Nuclear Spectroscopy and Center for Nuclear Matter Science, Institute of Modern Physics, Chinese Academy of Sciences, Lanzhou 730000, People's Republic of China}
\author{X.~H.~Zhou}
\affiliation{Key Laboratory of High Precision Nuclear Spectroscopy and Center for Nuclear Matter Science, Institute of Modern Physics, Chinese Academy of Sciences, Lanzhou 730000, People's Republic of China}
\author{F.~R.~Xu}
\affiliation{State Key Laboratory of Nuclear Physics and Technology, School of Physics, Peking University, Beijing 100871, People's Republic of China}
\author{G.~Audi}
\affiliation{CSNSM-IN2P3-CNRS, Universit\'{e} de Paris Sud, 91405 Orsay, France}
\author{R.~J.~Chen}
\affiliation{Key Laboratory of High Precision Nuclear Spectroscopy and Center for Nuclear Matter Science, Institute of Modern Physics, Chinese Academy of Sciences, Lanzhou 730000, People's Republic of China}
\author{X.~C.~Chen}
\affiliation{Key Laboratory of High Precision Nuclear Spectroscopy and Center for Nuclear Matter Science, Institute of Modern Physics, Chinese Academy of Sciences, Lanzhou 730000, People's Republic of China}
\author{C.~Y.~Fu}
\affiliation{Key Laboratory of High Precision Nuclear Spectroscopy and Center for Nuclear Matter Science, Institute of Modern Physics, Chinese Academy of Sciences, Lanzhou 730000, People's Republic of China}
\author{Z.~Ge}
\affiliation{Key Laboratory of High Precision Nuclear Spectroscopy and Center for Nuclear Matter Science, Institute of Modern Physics, Chinese Academy of Sciences, Lanzhou 730000, People's Republic of China}
\author{W.~J.~Huang}
\affiliation{Max-Planck-Institut f\"{u}r Kernphysik, 69117 Heidelberg, Germany}
\author{S.~Litvinov}
\affiliation{GSI Helmholtzzentrum f\"{u}r Schwerionenforschung, 64291 Darmstadt, Germany}
\author{D.~W.~Liu}
\affiliation{Key Laboratory of High Precision Nuclear Spectroscopy and Center for Nuclear Matter Science, Institute of Modern Physics, Chinese Academy of Sciences, Lanzhou 730000, People's Republic of China}
\author{Y.~H.~Lam}
\affiliation{Key Laboratory of High Precision Nuclear Spectroscopy and Center for Nuclear Matter Science, Institute of Modern Physics, Chinese Academy of Sciences, Lanzhou 730000, People's Republic of China}
\author{X.~W.~Ma}
\affiliation{Key Laboratory of High Precision Nuclear Spectroscopy and Center for Nuclear Matter Science, Institute of Modern Physics, Chinese Academy of Sciences, Lanzhou 730000, People's Republic of China}
\author{R.~S.~Mao}
\affiliation{Key Laboratory of High Precision Nuclear Spectroscopy and Center for Nuclear Matter Science, Institute of Modern Physics, Chinese Academy of Sciences, Lanzhou 730000, People's Republic of China}
\author{A.~Ozawa}
\affiliation{Insititute of Physics, University of Tsukuba, Ibaraki 305-8571, Japan}
\author{B.~H.~Sun}
\affiliation{School of Physics and Nuclear Energy Engineering, Beihang University, Beijing 100191, People's Republic of China}
\author{Y.~Sun}
\affiliation{Department of Physics and Astronomy, Shanghai Jiao Tong University, Shanghai 200240, People's Republic of China} 
\author{T.~Uesaka}
\affiliation{RIKEN Nishina Center, RIKEN, Saitama 351-0198, Japan}
\author{G.~Q.~Xiao}
\affiliation{Key Laboratory of High Precision Nuclear Spectroscopy and Center for Nuclear Matter Science, Institute of Modern Physics, Chinese Academy of Sciences, Lanzhou 730000, People's Republic of China}
\author{Y.~M.~Xing}
\affiliation{Key Laboratory of High Precision Nuclear Spectroscopy and Center for Nuclear Matter Science, Institute of Modern Physics, Chinese Academy of Sciences, Lanzhou 730000, People's Republic of China}
\author{T.~Yamaguchi}
\affiliation{Department of Physics, Saitama University, Saitama 338-8570, Japan}
\author{Y.~Yamaguchi}
\affiliation{RIKEN Nishina Center, RIKEN, Saitama 351-0198, Japan}
\author{X.~L.~Yan}
\affiliation{Key Laboratory of High Precision Nuclear Spectroscopy and Center for Nuclear Matter Science, Institute of Modern Physics, Chinese Academy of Sciences, Lanzhou 730000, People's Republic of China}
\author{Q.~Zeng}
\affiliation{Key Laboratory of High Precision Nuclear Spectroscopy and Center for Nuclear Matter Science, Institute of Modern Physics, Chinese Academy of Sciences, Lanzhou 730000, People's Republic of China}
\author{H.~W.~Zhao}
\affiliation{Key Laboratory of High Precision Nuclear Spectroscopy and Center for Nuclear Matter Science, Institute of Modern Physics, Chinese Academy of Sciences, Lanzhou 730000, People's Republic of China}
\author{T.~C.~Zhao}
\affiliation{Key Laboratory of High Precision Nuclear Spectroscopy and Center for Nuclear Matter Science, Institute of Modern Physics, Chinese Academy of Sciences, Lanzhou 730000, People's Republic of China}
\author{W.~Zhang}
\affiliation{Key Laboratory of High Precision Nuclear Spectroscopy and Center for Nuclear Matter Science, Institute of Modern Physics, Chinese Academy of Sciences, Lanzhou 730000, People's Republic of China}
\author{W.~L.~Zhan}
\affiliation{Key Laboratory of High Precision Nuclear Spectroscopy and Center for Nuclear Matter Science, Institute of Modern Physics, Chinese Academy of Sciences, Lanzhou 730000, People's Republic of China}



\date{\today}

\begin{abstract}
Isochronous mass spectrometry has been applied in the storage ring CSRe to measure the masses of the neutron-rich $^{\operatorname{52-54}}$Sc and $^{54,56}$Ti nuclei.
The new mass excess values $ME$($^{52}$Sc) $=$ $-40525(65)$ keV, $ME$($^{53}$Sc) $=$ $-38910(80)$ keV, and $ME$($^{54}$Sc) $=$ $-34485(360)$ keV,
deviate from the Atomic Mass Evaluation 2012 by 2.3$\sigma$, 2.8$\sigma$, and 1.7$\sigma$, respectively. 
These large deviations significantly change the systematics of the two-neutron separation energies of scandium isotopes.
The empirical shell gap extracted from our new experimental results shows a significant subshell closure at $N = 32$ in scandium, with a similar magnitude as in calcium.  
Moreover, we present $ab$ $initio$ calculations using the valence-space in-medium similarity renormalization group 
based on two- and three-nucleon interactions from chiral effective field theory. 
The theoretical results confirm the existence of a substantial $N = 32$ shell gap in Sc and Ca with a decreasing trend towards lighter isotones, 
thus providing a consistent picture of the evolution of the $N = 32$ magic number from the $pf$ into the $sd$ shell.
\end{abstract}


\maketitle

\section{Introduction}

The particularly bound and enhanced stable nature of some special nuclei with certain configurations of protons and neutrons 
led Mayer and Jensen to introduce the nuclear shell model~\cite{Mayer1,Mayer2}.
These are the well-known magic numbers associated with proton or neutron numbers 2, 8, 20, 28, 50, 82, and neutron number 126. 
In the single-particle shell model, protons and neutrons occupy nuclear orbitals with different quantum numbers. 
When the orbitals are fully filled, nuclides are much more bound than the neighboring ones.
The closed-orbit nuclei have typically spherical shapes. 
The magic numbers were established for nuclei close to the valley of $\beta$ stability.
However, the nuclear shell structure has been found to change when moving towards the drip lines.
For instance, a new shell gap at $N = 16$ has been observed establishing $^{24}$O as a doubly magic nucleus \cite{24O}.
The evolution of the nuclear shell structure at extreme proton-to-neutron ratios has become one of the key research quests \cite{shells}.

In the past decades, a lot of efforts have been made to study the shell evolution of $N = 32$ and 34 subshells, 
where protons ($\pi$) and neutrons ($\nu$) $p_{3/2}$-$p_{1/2}$ and $f_{7/2}$-$f_{5/2}$ spin-orbit partners determine the structure.  
A local maximum in the systematics of the first $2^+$ excitation energies [$E(2^+_1)$] in even-even nuclei at $N = 32$ were reported in 
$_{18}$Ar~\cite{50Ar}, $_{20}$Ca~\cite{52Ca1}, $_{22}$Ti~\cite{54Ti1}, and $_{24}$Cr~\cite{56Cr1} isotopes, remarkably suggesting a new neutron shell closure at $N = 32$.
Meanwhile, a local minimum in the systematics of reduced transition probabilities $B(E2; 0^+_1~\rightarrow~2^+_1)$ has also provided an evidence for the existence of this sub-shell in Ti~\cite{54Ti2} and Cr~\cite{56Cr2} isotopes. Furthermore, a sizable subshell closure with a similar magnitude as the $N = 32$ gap in $^{52}$Ca 
has been unambiguously demonstrated at $N = 34$ in $^{54}$Ca~\cite{54Ca}. 

The emergence and weakening of new subshell closures $N = 32$, 34 have been successfully elucidated within the shell model by the tensor force 
acting between protons in $j{}=l\pm \frac{1}{2}$ and neutrons in $j{'}{}=l{'}\pm \frac{1}{2}$ orbitals, 
where $l$ and $l{'}$ represent orbital angular momenta of protons and neutrons, respectively. 
In the standard shell model picture in this mass region, the valence protons in the ${\pi} f_{7/2}$ orbital 
have an attractive tensor force with the valence neutrons in the ${\nu} f_{5/2}$ orbital. 
As soon as the protons are removed from the ${\pi} f_{7/2}$ orbital, that is when going from $_{26}$Fe to $_{20}$Ca, 
the magnitude of the effect of the attractive $\pi$-$\nu$ tensor force decreases consequently resulting in an upshift of the ${\nu} f_{5/2}$ orbital. 
If the ${\pi} f_{7/2}$ is completely empty, a substantial energy gap may exist between 
the ${\nu} f_{5/2}$ orbital and ${\nu} p_{3/2}$-${\nu} p_{1/2}$ spin-orbit partners leading to the formation of a new subshell at $N = 34$.  
Furthermore, the spin-orbit splitting of the partners results in a sizable energy gap, the $N = 32$ subshell.  
The determination of the upper boundaries of these new sub-shells at $N = 32$, 34 provides information 
on the relative ordering of ${\nu} f_{5/2}$ and ${\nu} p_{3/2}$-${\nu} p_{1/2}$ spin-orbit partners and leads to 
a better understanding of the role of the tensor force on the shell evolution in exotic neutron-rich nuclei. 
The low-lying energy levels in $^{55}$Sc indicate a quite rapid reduction of the $N = 34$ sub-shell gap, 
even though only one proton is added to the ${\pi}f_{7/2}$ orbital~\cite{55Sc}.
The reduction of the $N = 34$ sub-shell gap in $^{56}$Ti~\cite{56Ti} and the robustness of the $N = 32$ sub-shell in $^{54}$Ti~\cite{54Ti1,54Ti2} 
reveals that the $\nu f_{5/2}$ orbital is still above the $\nu p_{3/2}$-$\nu p_{1/2}$ partners but is quite close to the $\nu p_{1/2}$ orbital 
if two protons are added (for more details see Fig. 1 and the related discussion in Ref.~\cite{54Ca}). 
We note that the shell model can well reproduce the experimental results~\cite{50Ar,54Ca}.

Due to the particularly strong binding nature of magic nuclei, the two-neutron separation energy,  $S_{2n}$, 
defined as $S_{2n} (Z,N) = BE(Z,N) - BE(Z,N-2)$, where $BE$ is nuclear binding energy, is a well-established signature of neutron shell gaps, when a sudden change in the slope of a smooth $S_{2n}$ systematics occurs.
The advantage of this indicator is that it is applicable not only in even-$Z$ isotopic chains but also in the odd-$Z$ ones. 
High-precision mass measurements~\cite{TITANPRL,ISOLDE} have confirmed the existence of the $N = 32$ sub-shell closure in calcium. 
Furthermore, mass measurements for $^{52,53}$K revealed the persistence of the $N = 32$  shell gap in potassium below the proton magic number $Z = 20$~\cite{52K}.
The overall picture is consistent with nuclear spectroscopy data mentioned above. 
However, the $S_{2n}$ behavior as a function of neutron number in the $N\approx32$ region becomes smooth for $_{22}$Ti~\cite{LEIS18}, $_{23}$V~\cite{55V}, and $_{24}$Cr~\cite{ISOCR}, indicating the reduction of the $N = 32$ subshell closure. 
 $Ab$ $initio$ calculations using the valence-space in-medium similarity renormalization group (VS-IMSRG) have successfully predicted binding energies of the nuclear ground states in this mass region \cite{SIM17,LEIS18,MOUG18}. While these calculations generally describe the overall trends pointing to new magic numbers, signatures such as $2^+$ energies, and neutron shell gaps tend to be modestly overpredicted, as seen recently  in the titanium isotopes~\cite{LEIS18}, 
 highlighting the need for future improvements in the many-body approach.

Due to large uncertainties (about several hundred keV) of nuclear masses for Sc isotopes around $A\approx 53$, 
it was difficult to give a definite conclusion on the $N = 32$ shell evolution above $Z = 20$,
or in other words, on the ordering of $\nu f_{5/2}$ and $\nu p_{3/2}$-$\nu p_{1/2}$ partners. In this paper, we report direct mass measurements of $^{\operatorname{52-54}}$Sc and address the question of the upper boundary of the $N = 32$ subshell closure. In addition, masses of $^{54,56}$Ti have been obtained.
After a very preliminary analysis reported in Ref.~\cite{CPC}, this work presents the final experimental results and their theoretical interpretation.

\begin{figure}[t]
	\includegraphics[width=1.0\columnwidth]{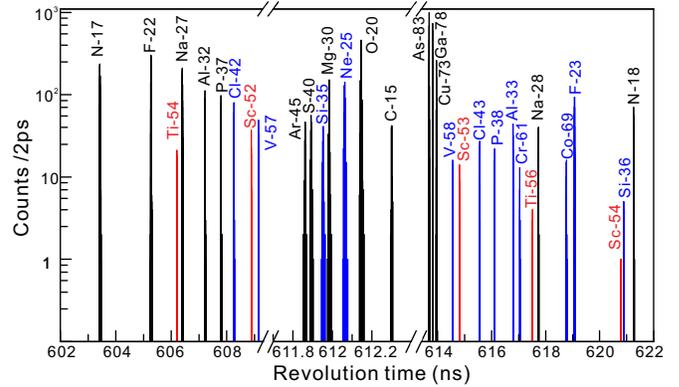}
	\caption{(Color online). Part of the measured revolution time spectrum
         in the time window 603 ns $\leq$ \emph{t}  $\leq 622$ ns. 
         {{The nuclides with well-known masses were used for calibration (black color).
         The nuclides whose masses were determined in this work are indicated with red color. 
         The determination of the masses of the remaining isotopes (blue color) is outside of the scope of the present paper.}}
}
  \label{spectrum}
\end{figure}

\section{Experiment}
The experiment was conducted at the Heavy Ion Research Facility in Lanzhou (HIRFL)~\cite{Xiajiawen}.
Primary $^{86}$Kr$^{28+}$ beams were accelerated to 460.65~MeV/u by the heavy-ion synchrotron CSRm.
They were fast-extracted and focused upon a $\approx $15~mm thick $^9$Be target placed in front 
of the in-flight fragment separator RIBLL2~\cite{zwl2010}. At this relativistic energy, the reaction products 
from the projectile fragmentation of $^{86}$Kr emerged the target predominantly as bare nuclei. 
They were analyzed~\cite{Geis92} by their magnetic rigidities $B\rho$ by the RIBLL2. 
A cocktail beam including the ions of interest was injected into the cooler storage ring (CSRe).
The isochronous mass spectrometry (IMS) technique \cite{Haus00,IMS,Hushan13,zyh2016a,2016XU10,Zhangp16,zhangyh18} 
has been applied in the CSRe for precision mass measurements of the stored ions.

The primary beam energy was selected according to the LISE++ simulations~\cite{Tar08} 
such that after the target the $^{61}$Cr$^{24+}$ ions had the most probable velocity with $\gamma$ $=$ $\gamma_t$ $=$ $1.40$, 
where $\gamma$ is the Lorentz factor and  $\gamma_t$ is the set CSRe transition point. 
For an optimal transmission of nuclides centered at $^{61}$Cr, RIBLL2 and CSRe were set to a fixed magnetic rigidity of $B\rho = 7.6755$~Tm.
The projectile fragments have a broad momentum distribution of a few percent, 
among which only those within the $B\rho$ acceptance of $\pm0.2\%$ of the RIBLL2-CSRe system 
have been transmitted and stored in the CSRe. 

The revolution times of the stored ions were measured with a time-of-flight (ToF) detector~\cite{Meibo} installed inside the CSRe aperture. 
At each revolution ions passed through a 19 ${\mu}$g/cm$^2$ carbon foil thereby releasing secondary electrons.
The latter were guided to a micro-channel plate (MCP) counter. The signals from the MCP were directly recorded by an oscilloscope.

The revolution frequencies of the ions were about 1.6 MHz. The resolution of the ToF detector was about 50 ps. 
For each injection, a measurement time of  $200$ $\mu$s, triggered by the CSRe injection kicker, 
was acquired, which corresponds to about $300$ revolutions of the ions in the CSRe. 
The efficiency varied from 20\% to 70\% depending the charge of ion species. Because only about five ions were stored simultaneously in each injection, the saturation effect of MCP did not occur. The typical efficiency for the nucleus of interest with ionic charge around 20 was about 50\%, see Refs.~\cite{Meibo,Tuxiaolin,Zhangwei} .
In total 10300 injections were accomplished. The revolution time spectrum and the corresponding isotope identification 
were obtained as described in Refs. \cite{Tuxiaolin,zhangyh18,Co51,TuPRL,ZhangPRL,magnetic}. A part of the measured spectrum is shown in Fig.~\ref{spectrum}.

Many of nuclides in Fig.~\ref{spectrum} have well-known masses. Their mass excess $(ME)$ values from AME$^{\prime}$12~\cite{AME2012} were used\footnote{%
Since our preliminary values from Ref. \cite{CPC} were included into the latest Atomic Mass Evaluation, AME$^\prime$16 \cite{AME16},
we use the values from the preceding AME$^\prime$12 \cite{AME2012} for comparison.}%
~to fit their mass-to-charge ratios $m/q$ versus the corresponding revolution times $T$ by employing a third-order polynomial function.
The mass calibration has been checked by redetermining the $ME$ values of each of the $N_c$ reference nuclides ($N_c = 15$) using the other $N_c~-~1$ ones as calibrants.
The normalized $\chi_{n}$ defined as:
\begin{eqnarray}\label{Chi-square equ}	\chi_{n}=\sqrt{\frac{1}{N_c}\sum\limits_{i=1}^{N_c}\frac{[(\frac{m}{q})_{i,\rm{exp}} - (\frac{m}{q})_{i,\rm{AME}}]^{2}}{\sigma^2_{{i,\rm{exp}}} + \sigma^2_{{i,\rm{AME}}}}}~,
\end{eqnarray}
was found to be $\chi_{n} = 0.97$. This value is within the expected range of $\chi_{n} = 1\pm0.18$ at $1\sigma$ confidence level, indicating that no additional systematic error has to be considered. 
\begin{figure}[hbp]
	\includegraphics[width=0.9\columnwidth]{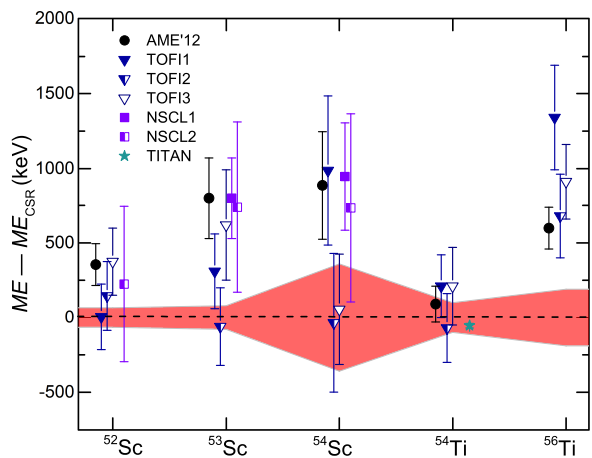}
	\caption{(Color online). Differences between $ME$ values determined in this work and other experiments (see legend, text, and Table~\ref{tablemass}). 
	The red shadings represent the uncertainties from this work while the error bars are uncertainties from other experiments.
}  \label{mutualcal}
\end{figure}

\section{results and discussion}
Figure~\ref{mutualcal} presents the differences between $ME$ values determined in this work and their previously known literature values. 
The obtained results are listed in Table \ref{tablemass}.
Owing to the large uncertainties of the ToF-$B\rho$ measurements established at a radioactive beam line, 
results of all experiments seem to be in general consistent at $3\sigma$ confidence level. 
The excellent agreement  between our results and the precision mass measurements from MR-ToF at TITAN for $^{54}$Ti \cite{LEIS18}, 
which is at the edge of the isochronous window \cite{Tuxiaolin}, proves the reliability of our measurements. 
All previous measurements were evaluated in the AME$^{\prime}$12 yielding recommended values, which are also illustrated in Fig.~\ref{mutualcal}.
It is striking that, except for the precision value of $^{54}$Ti, our new results significantly deviate from AME$^{\prime}$12 values,
namely by 2.3, 2.8, 1.7, and 2.5 standard deviations, respectively, for $^{52,53,54}$Sc and $^{56}$Ti nuclei.
We note that unpublished measurements from GSI \cite{MATOS} are in overall good agreement with our results. However, they were discarded in the AME$^\prime$12.

\begin{figure}[htb]
	\includegraphics[width=0.9\columnwidth]{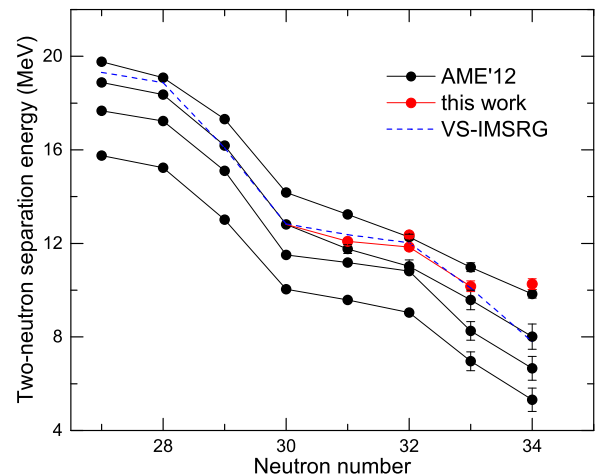}
	\caption{(Color online). $S_{2n}$ values for K, Ca, Sc, and Ti isotopic chains (see legend). 
	The remarkable agreement between the experimental data and VS-IMSRG calculations is clearly seen.}
	\label{s2n}
\end{figure}

Our new results completely change the systematic behavior of $S_{2n}$ of the scandium isotopic chain as a function of neutron number $N$.
As can be seen from Fig.~\ref{s2n}, the $S_{2n}$($^{52}$Sc) as well as $S_{2n}$($^{53}$Sc) are now significantly larger than assumed previously, 
and consequently, a kink at $N = 32$ emerges clearly. 
This behavior is in line with the recently established trends for calcium~\cite{ISOLDE} and potassium~\cite{52K} isotopic chains. 
Our results undoubtedly indicate the persistence of the sub-shell $N = 32$ in scandium.

\begin{table*}[htb]
	\caption{
	Mass excess $(ME)$ values in keV of $^{\operatorname{52-54}}$Sc and $^{54,56}$Ti from the present work, 
	three ToF-B$\rho$ measurements at TOFI-Los Alamos~\cite{TOFI1,TOFI2,TOFI3}, 
	two ToF-B$\rho$ measurements at NSCL-MSU~\cite{MSUPRL1,MSUPRL2}, and a MR-ToF measurement at TITAN-TRIUMF~\cite{LEIS18}.
    The $ME$ values from the  AME$^\prime$12~\cite{AME2012} and their deviations from our new results taking into account both error bars, 
		$\Delta/\delta=|{ME}_{\rm{CSRe}}-{ME}_{\rm{AME'12}}|/\sqrt{\delta({ME}_{\rm{CSRe}})^2+\delta({ME}_{\rm{AME'12}})^2}$, are given in the last two columns.
 }
\begin{tabular*}{\textwidth}{lccccccccc}
		\hline
		\hline
		Atom     & $ME_{\rm{CSRe}}$   &  $ME_{\rm{TOFI1}}$~\cite{TOFI1}  & $ME_{\rm{TOFI2}}$~\cite{TOFI2}     & $ME_{\rm{TOFI3}}$~\cite{TOFI3} & $ME_{\rm{NSCL1}}$~\cite{MSUPRL1}    & $ME_{\rm{NSCL2}}$~\cite{MSUPRL2}      &         $ME_{\rm{TITAN}}$~\cite{LEIS18}         & $ME_{\rm{AME^{\prime}12}}$~\cite{AME2012}&    $\Delta/\delta$  \\
$^{52}$Sc     & $-40525(65)$   &  ~$-40520(220)$~    &~ $-40380(230)$~     & ~$-40150(225)$~  &            -               & $-40300(520)$ &           -                    & $-40170(140)$  & 2.31\\
$^{53}$Sc     & $-38910(80)$ &   ~$-38600(250)$~     & ~$-38970(260)$~     & ~$-38290(370)$~  & $-38110(270)$ & $-38170(570)$ &            -                      &  $-38110(270)$ &2.84\\
$^{54}$Sc     & $-34485(360)$ &  ~$-33500(500)$~     & ~$-34520(465)$~    & ~$-34430(370)$~   & $-33540(360)$ & $-33750(630)$ &              -                   &  $-33600(360)$ & 1.74\\
$^{54}$Ti     & $-45690(100)$ &    ~$-45480(210)$~      & ~$-45760(230)$~    &~$-45480(260)$~    &           -                  &          -                   &$-45744(16)$  &  $-45600(120)$  &0.58\\
$^{56}$Ti     & $-39810(190)$ &  ~$-38470(350)$~      & ~$-39130(280)$~    & ~$-38900(250)$~   &              -               &             -                &         -                &  $-39210(140)$  & 2.54\\
		\hline
	\end{tabular*}
	\label{tablemass}
\end{table*}

\begin{figure}[htb]
	\includegraphics[width=0.9\columnwidth]{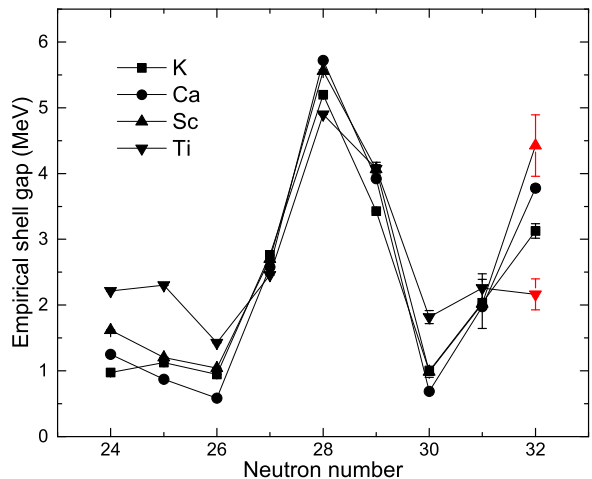}
	\caption{(Color online). Empirical shell gap for K, Ca, Sc, and Ti isotopic chains. 
	The shell gap at $N = 28$ is nicely seen in all four elements. 
	The shell gap at $N = 32$ is well pronounced in scandium and calcium and is strongly reduced in titanium.\\
	}
	\label{shellgap}
\end{figure}

The strength of neutron subshell/shell closures can be evaluated via the empirical neutron shell gap energy, 
defined as the difference of two-neutron separation energies $\Delta_{2n}(N,Z) = S_{2n}(N,Z)$ $-$ $S_{2n}(N+2,Z)$. 
As seen in Figure~\ref{shellgap}, the $N = 32$ gap is enhanced in calcium and scandium up to values 
comparable to that of the well-known $N = 28$ shell gap, suggesting the robustness of a prominent $N = 32$ subshell closure. 
We emphasize that a rapid reduction of the $N = 32$ shell gap is confirmed experimentally in titanium \cite{LEIS18} and beyond, 
where two or more protons occupy the $\pi f_{7/2}$ orbital. 

In VS-IMSRG calculations, we have found a strong reduction in the $N = 32$ shell gap only at the vanadium isotopes, 
overpredicting the shell gap in titanium compared to experiment. 
In this work, we have performed calculations for the scandium, calcium, and potassium isotopes to determine the evolution of the shell gap across $Z = 20$.
In particular, we use a VS-IMSRG approach \cite{TSU12,BOG14,STR16}, where an approximate unitary transformation \cite{MOR15,HER16} is constructed to first decouple the $^{40}$Ca core, as well as a standard $pf$ valence-space Hamiltonian. This interaction is subsequently diagonalized using the {\sc nushellx}@{\sc msu} shell-model code \cite{BRO14}. 
We further capture the effects of three-nucleon ($3N$) forces between valence nucleons through the ensemble normal ordering \cite{STR17}, which gives a unique valence-space Hamiltonian for each nucleus.
We are then able to test nuclear forces in essentially all open-shell systems accessible to the nuclear shell model 
with a level of accuracy comparable to large-space $ab$ $initio$ methods~\cite{STR17}. 

\begin{figure}[htb]
	\includegraphics[width=0.85\columnwidth]{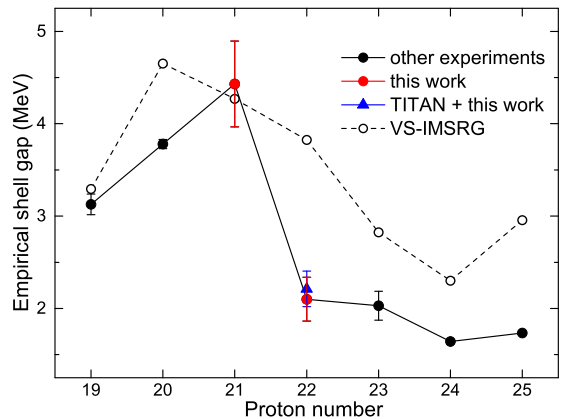}
	\caption{(Color online). Empirical shell gap values for $N = 32$ isotones. Black points are obtained from AME$^\prime$12~\cite{AME2012} while color points are from recent mass measurements by different labs. A significant shell gap in scandium is well reproduced by the theory as well as the decrease of the shell gap towards heavier elements.\\
	}
	\label{shellevolution}
\end{figure}

We use the EM(1.8/2.0) $NN + 3N$ interactions of Refs.~\cite{HEB11,SIM16}, 
which begins from the chiral $NN$~N$^3$LO potential of Ref.~\cite{EN03} combined with a non-local $3N$ force fit in $A = 3$, 4-body systems, 
but which reproduces ground- and excited-state energies to the tin region and beyond \cite{SIM17,MOR18}. 
Predictions from this approach agree well with new experimental ground-state energy measurements in nearby titanium \cite{LEIS18} and chromium \cite{MOUG18} isotopes, 
as well as spectroscopy in neutron-rich scandium \cite{GARN17,55Sc}. 

We first note that, while not shown, absolute ground-state energies of neutron-rich scandium isotopes are well reproduced, generally at the level of $1$-$2\%$ or better. 
As seen in Fig.~\ref{s2n}, $S_{2n}$ values are also very well reproduced along the chain, 
including the sharp drops in $S_{2n}$ at $N = 28$ and $N = 32$ and the deviation from AME$^\prime$12 values at $N = 31$, 33. Figure~\ref{shellevolution} illustrates the experimental and calculated empirical shell gap for $N = 32$ isotones. 
The theoretical values describe reasonably the trend of experimentally determined shell gaps from potassium to manganese, including the sharp peaks at calcium and scandium. However, we note that in general the shell gaps are overpredicted by several hundred keV, but are particularly high in titanium and calcium, as first noted in Ref.~\cite{LEIS18}. 
The origin of this deviation is not yet fully understood, but signatures of shell closures are often modestly overestimated by the current level of many-body truncations implemented in the VS-IMSRG  \cite{SIM17}. This overprediction is also consistently seen in first excited $2^+$ energies, where predictions are 300-400 keV too high in both $^{52}$Ca and $^{54}$Ti. From benchmarks with coupled-cluster theory \cite{MOR18}, it is expected that improved treatments for currently neglected three-body operators in the VS-IMSRG will improve these predictions.

\section{Summary and Conclusions}

In summary, the masses of $^{\operatorname{52-54}}$Sc and $^{54,56}$Ti nuclides have been directly measured in the heavy ion storage ring CSRe in Lanzhou 
by employing isochronous mass spectrometry.
With the new mass values the previously known mass surface has been significantly modified. The existence of a strong $N = 32$ shell gap in scandium is evident. 
According to our experimental results the $N = 32$ shell gap is the largest in scandium. Furthermore, our new data confirm the absence of a significant shell gap in titanium.
The $ab$ $initio$ calculations using the VS-IMSRG approach with $NN$ and $3N$ interactions from chiral effective field theory confirm the experimental observations for calcium and scandium, but predict a persistence of a large $N = 32$ gap in titanium, at odds with these and other experimental measurements. Work is currently underway to improve the IMSRG approach to include the physics of neglected three-body operators, likely the origin of this discrepancy.  
The understanding of shell closures in neutron-rich nuclei is not only important for nuclear structure 
but is also critical for the reliable modeling of the structure of compact stellar objects as well as for nucleosynthesis.

\acknowledgments 
We thank the staffs of the accelerator division of the IMP for providing stable beam. 
This work is supported in part by the National Key  R\&D Program of China (Grant No. 2018YFA0404401 and No. 2016YFA0400504),
the NSFC (Grants No. 11605249, No. 11605248, No. 11605252, No. 11505267, No. 11575112, and No. 11575007), 
the CAS External Cooperation Program (Grant No. GJHZ1305), 
the CAS "Light of West China" Program, 
the CAS visiting professorship for senior international scientists (Grant No. 2009J2-23),
the CAS through the Key Research Program of Frontier Sciences (Grant No. QYZDJ-SSW-SLH005), 
the Helmholtz-CAS Joint Research Group (HCJRG-108),
the HGF Nuclear Astrophysics Virtual Institute (NAVI),
the Deutsche Forschungsgemeinschaft (DFG, German Research Foundation) -- Projektnummer 279384907 -- SFB 1245,
the BMBF (Contract No. 05P18RDFN1),
the DAAD PPP program with China (Project-ID 57389367),
the National Research Council of Canada and NSERC,
the DFG through the Cluster of Excellence PRISMA,
and the European Research Council (ERC) (ERC-StG 307986 "STRONGINT" and ERC-CG 682841 "ASTRUm"),
A.S. is supported by the Max Planck Society.
X.L.T. acknowledges the support from the Max Planck Society through the Max-Planck Partner Group.
Furthermore computations were performed with an allocation of computing resources at the J\"ulich Supercomputing Center (JURECA).

\end{document}